\documentclass [11pt]{article}
\title{Is detection of Fitzgerald-Lorentz contraction possible?}
\author{by Robert D. Klauber\\1100 University Manor Dr., 38B \\Fairfield, Iowa 52556
\\rklauber*AT*iowatelecom.net, permanent: rklauber*AT*netscape.net}

\date{August 7, 2008}

\usepackage{graphicx}
\usepackage{longtable}

\begin{document}

\maketitle
\begin{abstract}

Visual perception of Fitzgerald-Lorentz contraction is known to be 
theoretically impossible, and this can be demonstrated pedagogically with 
the aid of simple spacetime diagrams of one spatial dimension. Such diagrams 
also demonstrate, simply and directly, that the apparent length of a moving 
meter stick changes as it passes by and can even look elongated. In 
addition, measurement of a moving meter stick with instruments, as opposed 
to visual perception, must be inherently ambiguous, as the length measured 
depends on clock synchronization, which is widely considered to be 
conventional. In fact, for some synchronization choices, a moving meter 
stick would be measured as greater than one meter. Thus, the well known 
Fitzgerald-Lorentz contraction factor $\sqrt {1-v^2/c^2} $ would generally 
not be seen visually, and would only be measured in a system employing 
one particular (Einstein) synchronization convention.

\end{abstract}

\section{Introduction}
\label{sec:introduction}
\subsection{Background}
Although statements such as ``.. the experimental evidence for relativity is 
so overwhelming that physicists now regard {\ldots} [length contraction and 
time dilation] as commonplace'', (Kleppner and 
Kolenkow\cite{Kleppner:1973}), and `` {\ldots} special relativity theory 
represents a theory experimentally checked in all of its aspects'' 
(Brumberg\cite{Brumberg:2005}) are widespread, they are not 100{\%} true.

Certainly, time dilation, mass-energy equivalence, and the invariance of two 
way light speed were verified early on by the 
Michelson-Morley,\cite{Michelson:1} Kennedy-Thorndike,\cite{Kennedy:1932} 
Ives-Stilwell,\cite{Ives:1941} and various cyclotron experiments. Many 
later experiments greatly improved on the accuracy of the earlier ones (see 
Haugen and Will,\cite{Haugan:1987} and Klauber.\cite{Klauber:2004})

Somewhat obscured in summaries of experiments testing special relativity 
theory (SRT) is the fact that Fitzgerald\cite{now:1}-Lorentz contraction of 
moving bodies has never been directly observed. The reason often cited is 
one of practicality. At low speeds, the effect is simply too small to 
detect. At high speeds, the effect may be greater, but monitoring length on 
an object zipping by at close to the speed of light is well beyond the 
capability of real-world instruments.

One might therefore pose a reasonable question. If instruments of sufficient 
accuracy could one day exist, could we then witness Fitzgerald-Lorentz 
contraction? The answer reveals some intriguing facets of SRT.

\subsection{Seeing vs. Measuring: Framing the Question}
In SRT we often refer to an ``observer'', and concomitant with that term is 
the tendency to think of someone who ``observes'' (i.e., ``sees'' with the 
eyes) such things as moving rods looking shortened. Unfortunately, the term 
is misleading.

What we are really referring to in SRT when we speak of an observer, is a 
``measurer''. The equations of SRT incorporate quantities an experimentalist 
would measure at a given event. If we are distanced from certain events, 
what we see are the light rays emanating from them. These do not reach our 
eyes instantaneously, of course, but travel to us at the speed of light. So, 
light from two events which occur simultaneously, but are located at 
different distances from an observer, will be seen at different times by 
that observer. So from the point of view of ``seeing'' the events, they do 
not appear simultaneous, though from the point of view of ``measuring'' the 
events with clocks at their respective locations, they are. As first noted 
by Lampa\cite{Lampa:1924} and Terrell\cite{Terrell:1959}, this has 
significant implications for Fitzgerald-Lorentz contraction.

So we have to concern ourselves with two different questions.

\begin{enumerate}
\item Could we ever \textit{see} Fitzgerald-Lorentz contraction?
\item Could we ever\textit{ measure} it?
\end{enumerate}
After a brief review of Fitzgerald-Lorentz contraction in SRT in Section 
\ref{sec:fitzgerald}, we will address the first question in Section 3, 
the second in Section 4, related issues in Sections 
\ref{sec:revisiting} and \ref{sec:mylabel1}, and for 
completeness, concomitant issues regarding time dilation observation in 
Section \ref{sec:mylabel3}. The central theme of Section 
3 is known, though certainly not widely, and has 
not before, I submit, been presented in as simple and pedagogic a manner as 
it is herein.

\section{Fitzgerald-Lorentz Contraction Reviewed}
\label{sec:fitzgerald}
The contraction of length of moving objects is traditionally deduced from 
the Lorentz transformation,
\begin{equation}
\label{eq1}
\begin{array}{l}
 c{T}'=\gamma (cT-\frac{v}{c}X)\,\,\,\,\,\,\,\,\,\,\,\,\,\,\,(a) \\ 
 {X}'=\gamma (X-\frac{v}{c}cT)\,,\,\,\,\,\,\,\,\,\,\,\,\,\,\,\,(b) \\ 
 \end{array}
\end{equation}
where two spatial dimensions are suppressed, primed and non-prime 
coordinates represent two different coordinate frames with common origin at 
$T$ = 0, $v=\vert $\textbf{v$\vert $} is the speed of either frame measured in 
the other, spatial axes are collinear and aligned with \textbf{v}, $c$ is the 
speed of light, and$\gamma =\frac{1}{\sqrt {1-v^2/c^2} }$. The length of a 
moving rod is determined as the distance between endpoints of the rod 
\textit{provided the determination is made when the endpoints are at the same moment in time} (i.e., they are simultaneous in the observer/measurer's coordinate system.)

Thus, we consider two events A and B which occur simultaneously in the $X-T$ 
coordinate system and are located at opposite ends of a meter stick fixed in 
the ${X}'-{T}'$ system. From (\ref{eq1})(b),
\begin{equation}
\label{eq2}
{X}'_B -{X}'_A =\gamma \left( {X_B -X_A -\frac{v}{c}\underbrace {(cT_B -cT_A 
)}_{=\,\,0}} \right),
\end{equation}
where the time terms drop out because of simultaneity of A and B ($T_{A}$ = 
$T_{B}$.) Thus a distance of one meter measured as ${X}'_B -{X}'_A $ in the 
primed frame is less than one meter by 1/$\gamma $ as measured in the 
unprimed frame.

This is shown graphically in Fig. 1, where, as is well known, orthogonality in Minkowski space comprises space and time axes with inverse slopes.  Note in the figure that the meter stick fixed
in the primed frame has length as measured therein, with simultaneous 
endpoints (${T}'_C ={T}'_A )$, as ${X}'_C -{X}'_A $ = 1 m, and since ${X}'_C 
={X}'_B $, then ${X}'_B -{X}'_A $ also equals unity in (\ref{eq2}).

\begin{figure}
\centerline{\includegraphics{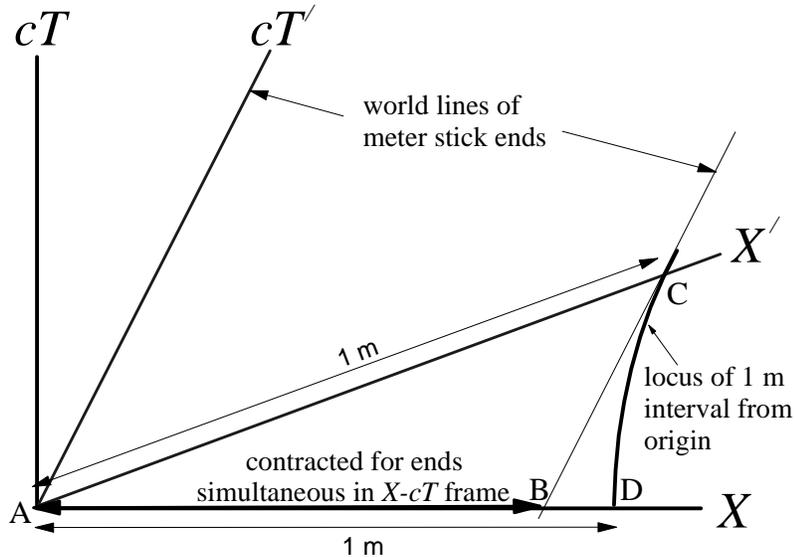}}
\caption{Contraction of a Moving Meter Stick in SRT}
\label{fig1}
\end{figure}

We, in the unprimed frame, have grid lines at one meter intervals laid out 
along our X axis. Events A and D are on two of these grid lines, one meter 
apart. We record the location at which each endpoint of the moving meter 
stick is passing (events A and B) when the clock at each location has the 
same reading ($T$ = 0 in Fig. 1). The distance between these locations is less 
than 1 meter (i.e., $X_{B}$ -- $X_{A} <$ 1m), so we consider the moving 
meter stick to be contracted.

In the unprimed frame, the simultaneity events at the endpoints are 
different events than they are in the primed frame, and this results in the 
famed Fitzgerald-Lorentz contraction.

\section{Visual Images of Moving Meter Sticks and Fitzgerald-Lorentz Contraction}
\label{sec:visual}
The result of the previous section has everything to do with measurement, 
particularly the measurement of the local clock readings at events A and B, 
and nothing to do with what an observer would see with her eyes.

\subsection{Relativistic ``Seeing'' Historically}
In a seminal, but little appreciated article, Lampa\cite{Lampa:1924} 
first addressed the issue of visual appearance of a relativistically moving 
rod. In further ground breaking work, Terrell\cite{Terrell:1959} showed 
that a three dimensional object traveling at relativistic speed would appear 
in a photograph as rotated and changed in scale. Many authors subsequently 
expanded on Terrell's work. Among these, Weinstein\cite{Weinstein:1960}, 
Scott and Viner\cite{Scott:1965}, Hickey\cite{Hickey:1979}, and 
Deissler\cite{Deissler:2005} noted the particular phenomenon addressed in 
the present Section 3, though none, I contend, have 
done so in a manner as pedagogically suited for presentation to new students 
of relativity. Specifically, we take a simpler tack and consider only one 
dimensional moving meter sticks, portrayed in spacetime diagrams of only one 
spatial dimension, such as that of Fig. 1. 

\subsection{Defining ``Seeing''}
\label{subsec:defining}
By ``see'' herein, we mean the following. A stationary observer close to the 
line of travel of a moving meter stick looks at one end of that meter stick 
and sees which grid line in her coordinate system that end looks coincident 
with. By doing likewise with the other end at the same moment (on her local 
clock), she can count the grid lines between endpoints and thus determine 
the length she is seeing. This eliminates the need to correct for apparent 
diminution, visually, of objects that are more distant, as well as the need 
for considering more than one dimension for viewing perspective. Further, it 
parallels the method we have for measuring any length or velocity 
non-locally. That is, we consider length or velocity to be that which would 
be measured by meter sticks proximate to the object being measured.

The following analysis is ideal, however, in that, being one dimensional 
(spatially), it assumes the viewer is coincident with the line of travel of 
the moving meter stick. In the real world, the observer would then only be 
able to see the front face of the meter stick and unable to see which grid 
lines the ends are adjacent to. Fuller, and more complicated, analyses by 
Scott and Viner\cite{Scott:1965}, Hickey\cite{Hickey:1979}, and 
Deissler\cite{Deissler:2005} show that a viewer close to, but not 
precisely aligned with, the line of travel will see, to high approximation, 
the meter stick behavior described in the simplified, single spatial 
dimension case considered herein.

\subsection{Relativistic Seeing Via Spacetime Diagrams}
As can be gleaned from Fig. 2, due to the requirement that light photons 
must travel to an observer's retina, and this takes time, two observers 
(O$_{1}$ and O$_{2})$ fixed at different locations in the unprimed frame 
will not actually see the same length for a meter stick moving relative to 
them. Fig. 2 shows the paths taken by light rays from each end of the moving 
meter stick that both reach an observer's eyes at the same time (same event) 
and demonstrates that each observer will generally \textit{not} see a moving meter stick 
contracted by 1/$\gamma $.

\begin{figure}
\centerline{\includegraphics{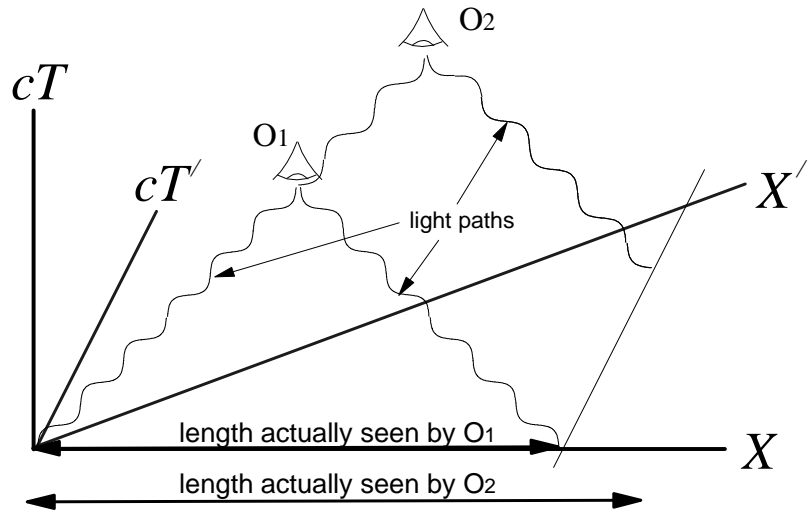}}
\caption{Contraction Seen Varies with Observer Position}
\label{fig2}
\end{figure}

As Gamba\cite{Gamba:1996} noted, ``No one will ever see the Lorentz 
contraction. [To be able to do so], one has to assume an infinite velocity 
of light{\ldots}''. Fitzgerald-Lorentz contraction is not something an 
observer simply sees visually.

\begin{figure}
\centerline{\includegraphics{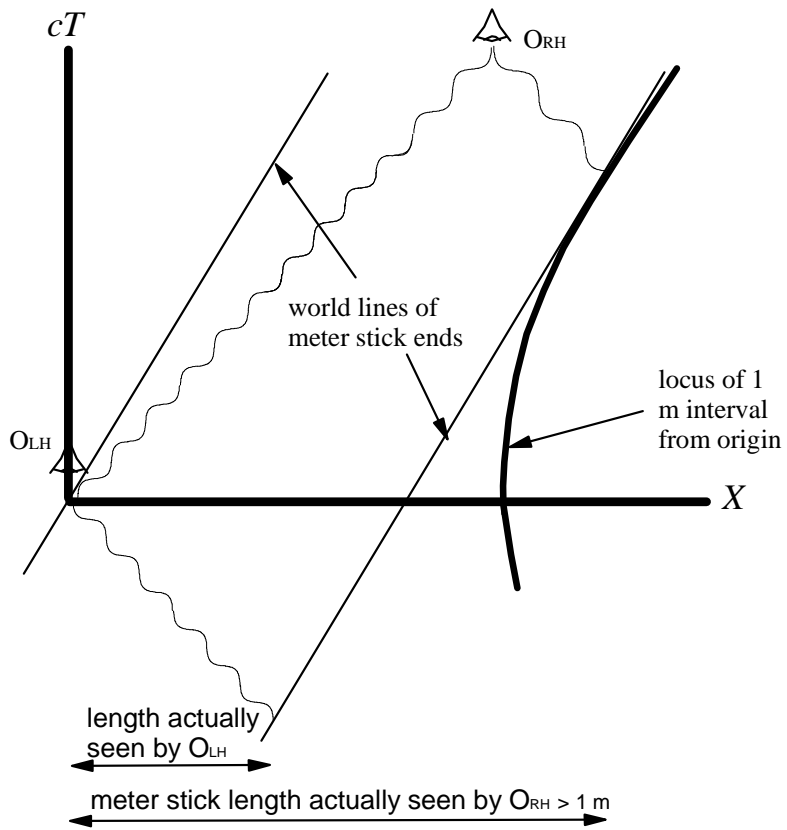}}
\caption{Elongation Seen by Some Observers}
\label{fig3}
\end{figure}
\begin{figure}
\centerline{\includegraphics{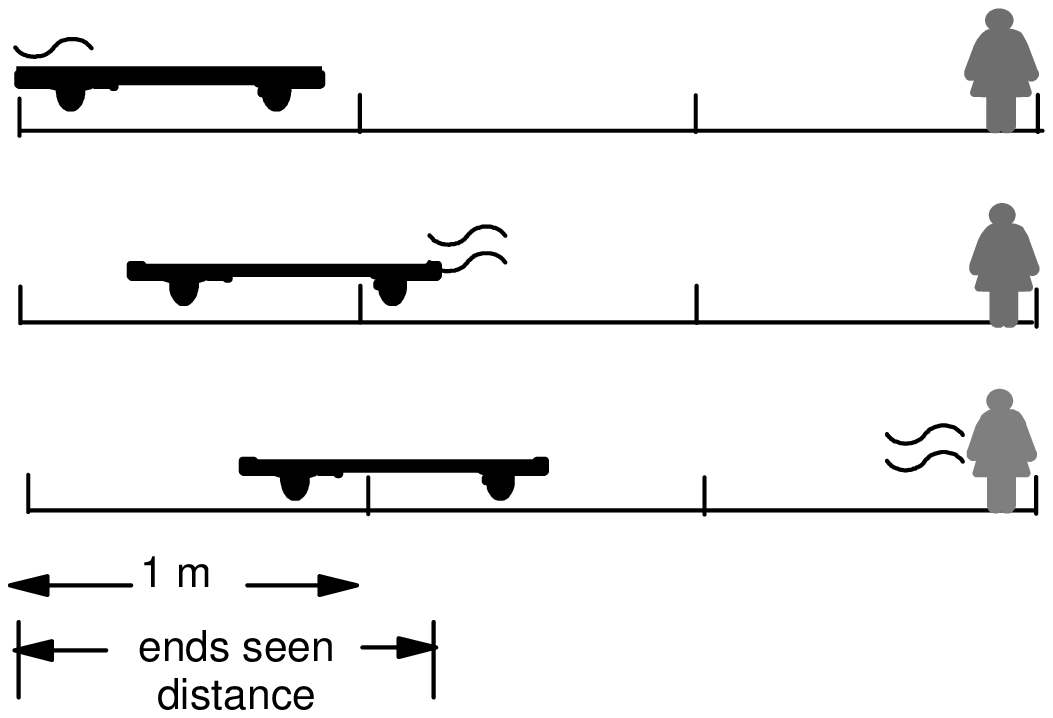}}
\caption{How a Moving Meter Stick Looks Elongated to an Observer in 
Front}
\label{fig4}
\end{figure}

It should also be noted that the length seen changes as the object moves 
past the seer. This can be understood by making other sketches similar to 
Fig. 2 in which one of the observers is shown at a different time (same $X$, 
different $T$.). It can also be understood from Fig. 3 by simply noting that an 
observer (O$_{RH})$ nearer the leading end of the moving object sees a 
longer length than an observer (O$_{LH})$ near the trailing end, and that as 
the object moves by, the relative positioning of an observer to the moving 
object changes.

Surprisingly, for certain observers at certain times, such as O$_{RH}$ in 
Fig. 3, the moving meter stick length can not only look greater than the 
Lorentz contracted length, but greater than one meter. 

Fig. 4 is physical depiction of this, showing a ``stationary'' observer 
situated ahead of a meter stick mounted on a moving train. Light from the 
trailing end of the moving meter stick must be released before light from 
the leading edge, in order for both to arrive at the observer at the same 
instant. But during the time between departures of the two light pulses, the 
meter stick moves. Hence, it appears longer than one meter to the observer 
shown in the figure.

With a modicum of thought, one can convince oneself that an approaching, 
non-accelerating meter stick looks longer than one meter by a constant 
value, until the moment we see the leading edge coincident with us. Then the 
meter stick starts looking shorter and shorter, for an instant (when we see 
ourselves as midway between the ends) looks contracted by the Lorentz 
factor, and then looks like it continues to shrink, until the trailing edge 
passes by us. Thereafter, the moving meter stick, looking more contracted 
than 1/$\gamma $ m, continues to look this same constant length as it 
recedes.

\subsection{Quantifying visual perception}
We can use Fig. 5 to determine the precise length $l_{a}$ we would see for an 
approaching meter stick having speed $v$. A stationary observer at event M, 
when she is coincident with the meter stick leading edge, sees light from 
the trailing edge that left at event N. The length such observer would see 
would equal the spatial distance between events Q and M. To find this, we 
need to find the distance $\vert X_{N}\vert $, the magnitude of the 
spatial coordinate of event N, and add that to $\sqrt {1-v^2/c^2} $, the 
spatial coordinate of event M. To do this, we need only to find the 
intersection of two world lines, that of the light ray and that of the 
trailing edge.

The trailing edge world line is
\begin{equation}
\label{eq3}
cT=\frac{c}{v}X,
\end{equation}
and the light ray world line is
\begin{equation}
\label{eq4}
cT=X-\sqrt {1-v^2/c^2} .
\end{equation}
Solving simultaneously, we find
\begin{equation}
\label{eq5}
X_N =-\frac{v}{c}\frac{\sqrt {1-v^2/c^2} }{1-v/c},
\end{equation}
and thus

\begin{figure}
\centerline{\includegraphics{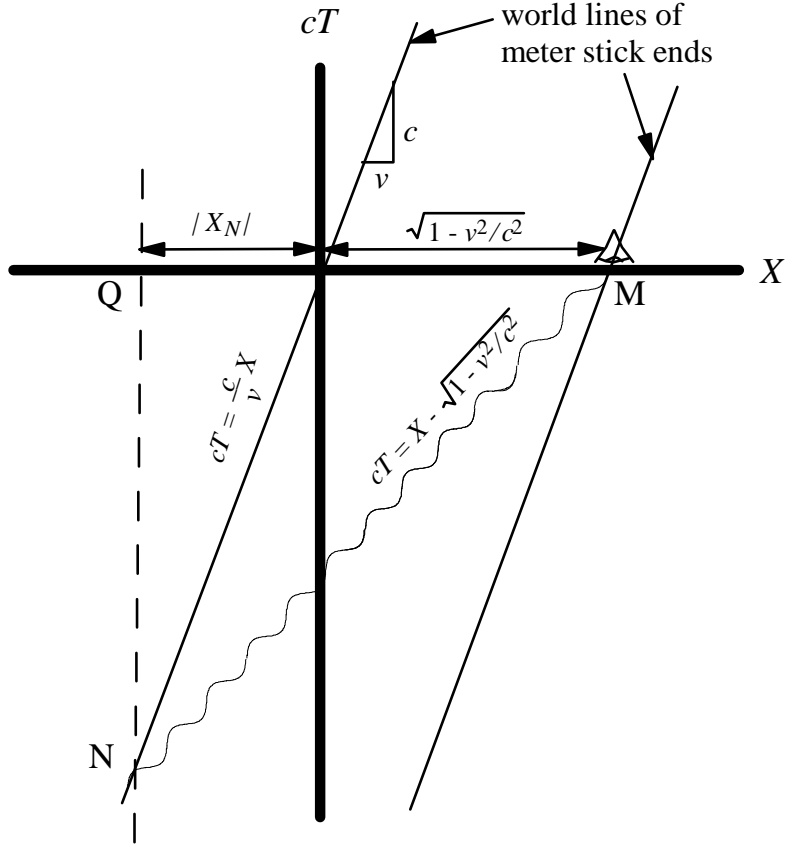}}
\caption{Determining Apparent Length of Approaching Meter Stick}
\label{fig5}
\end{figure}
\begin{equation}
\label{eq6}
l_a =\left| {X_N } \right|+\sqrt {1-v^2/c^2} =\frac{\sqrt {1-v^2/c^2} 
}{1-v/c}.
\end{equation}
In similar fashion, one can derive the comparable relation for the length 
$l_{r}$ of a receding meter stick,
\begin{equation}
\label{eq7}
l_r =\frac{\sqrt {1-v^2/c^2} }{1+v/c}\quad ,
\end{equation}
which differs from (\ref{eq6}) only in the sign before the speed term in the 
denominator. These relations are plotted in Fig. 6.

\subsection{Conclusion: Visual Perception of Length}
\label{subsec:conclusion}
We conclude that Fitzgerald-Lorentz contraction is not something that is 
actually seen by physical observers, and that the length change seen varies 
with observer position and as the observed object moves.

\begin{figure}
\centerline{\includegraphics{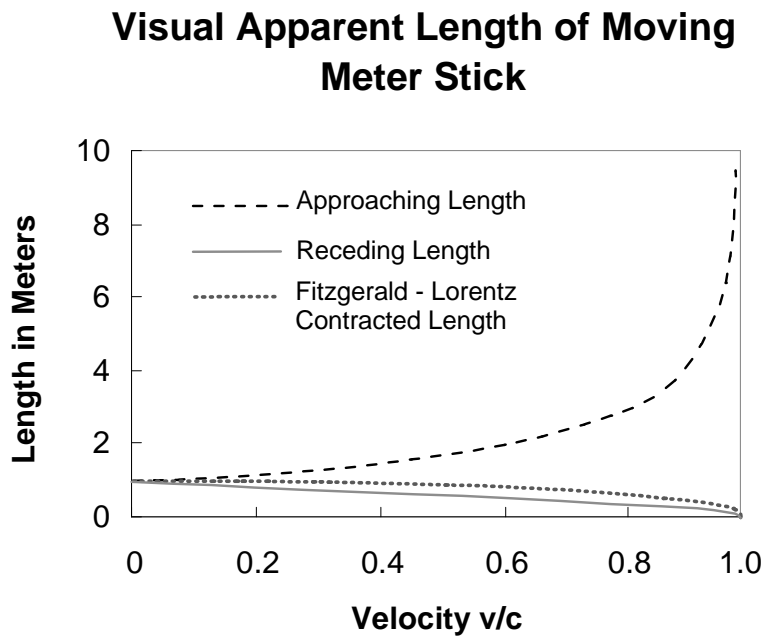}}
\caption{Apparent Meter Stick Length vs. Speed}

\label{fig6}
\end{figure}

\section{Fitzgerald-Lorentz Contraction and Conventionality of Synchronization}
\label{sec:mylabel2}
One might then ask if Fitzgerald-Lorentz contraction is, in theory, readily 
\textit{measurable}, even though it cannot be \textit{visually seen}. The answer to this question is both subtle and 
interesting, and involves a contemporary issue in relativity theory, 
conventionality of synchronization.

\subsection{Conventionality of Synchronization}
Synchronization in SRT involves the setting of standard clocks at different 
locations within a given reference frame. Since any number of settings on a 
given clock are possible, one must decide on a convention to use. The most 
common of these is the Einstein synchronization convention.

In Einstein synchronization, the one-way speed of light is assumed 
invariant, isotropic, and equal to $c$. One starts with a clock at one's own 
location, and at time $t_{1}$ on that clock, sends a pulse of light to a 
second location, whose distance $d$ from the first is known. The time elapsed 
during the flight of the light pulse is $d/c$, so one sets the time on the second 
location clock such that at the instant it received the light flash, it 
would have read
\begin{equation}
\label{eq8}
t_2 =t_1 +d/c.
\end{equation}
Note this ensures that any one-way measurement of light speed will result in 
a value of $c$, given that we use the local clocks at the emission and 
reception points to measure $\Delta t=t_2 -t_1 $.

Note also that the two-way (back and forth, reflected) speed of light is 
measured using only the clock at the first location, so no synchronization 
convention is needed to do that. The invariance of two-way light speed has 
been measured in many experiments, beginning with Michelson-Morley, and, for 
inertial systems at the least, has always been found invariant and equal to 
$c$.

However, as has been argued by many (see references in Anderson et 
al\cite{Anderson:1998}), there is an interrelationship between one-way 
light speed and synchronization convention. Synchronize your clocks in a 
different way, and the one-way speed of light is different (and not equal to 
$c$.) Conversely, assume a different one-way speed of light, and you get a 
different synchronization scheme for the clocks in your coordinate system.

The class of acceptable alternative synchronization schemes is limited to 
those for which the two-way speed of light remains invariant and equal to 
$c$, as has been proven experimentally. That is, the one-way speed of light in 
one direction would be less than $c$; in the other direction, greater than $c$; 
and the dependence would be such that the average round trip speed remains 
$c$.

Much work has been done on this, and most who have done that work believe 
that non-Einstein synchronizations, though more cumbersome mathematically, 
predict the very same observable quantities as traditional Einstein synched 
SRT. That is, there is no way to discern, via experiment, between 
synchronizations. They are merely different conventions, any one of which 
can be considered a valid representation of the physical world, and in this 
sense, synchronization is merely a gauge. (See Ohanian,\cite{Ohanian:2005} 
Choy,\cite{Choy:2004} Martinez,\cite{Martinez:2005} 
Macdonald,\cite{Macdonald:2005} and Klauber\cite{Klauber:2004b} for a debate over the opposing position.)

\subsection{Synchronization Convention and Fitzgerald-Lorentz contraction}
\label{subsec:synchronization}
Fig. 7 is a graphical illustration of the effect of an alternative 
synchronization in the stationary frame. Note that since simultaneity means 
having the same time on two different clocks, for the \textit{X-cT} system, the $X$ axis 
represents events which are all simultaneous (all with $T$ = 0). This is an 
Einstein synched coordinate system, which results in horizontal space and 
vertical time axes. For an alternative synchronization convention (\textit{x-ct} 
coordinate system with dashed coordinate grid lines) in the same frame, 
different events are simultaneous, so the set of such events with $t$ = 0 (the 
$x$ axis) is no longer horizontal (though the time axis remains vertical.)

Now, consider deducing the contraction effect for two different 
synchronization schemes. We use the same criterion as commonly used in 
traditional SRT, i.e., the endpoint events of the moving rod are measured 
when they are simultaneous.

As can be seen in Fig. 7, each synchronization convention (spatial axis 
slope) results in a different measured length. That is, the intersection 
point of the moving meter stick's right end worldline with the spatial axis 
will be different for different simultaneities in the stationary frame, and 
thus the length seen in that frame would depend on the simultaneity chosen 
therein. 

So if synchronization is a convention, then measurement of the contraction 
of a moving rod is also conventional\cite{Ref:1} \cite{Anderson:1992}. 
That is, we are hard pressed to say that there really is an absolute 
physical effect, agreed to by everyone and witnessed in the real world, 
whereby moving objects contract by a particular factor 1/$\gamma $. 

Fig. 8 demonstrates another surprising feature of conventionality of 
synchronization. That is, not only can the \textit{measured} length of a moving meter stick 
be greater than 1/$\gamma $ m, it can, for some synchronization choices 
(such as the \textit{x-ct} coordinates in Fig. 8), actually be greater than 1 m. The 
moving meter stick does not have to be measured as contracted, it can be 
\textit{elongated}.

\begin{figure}
\centerline{\includegraphics{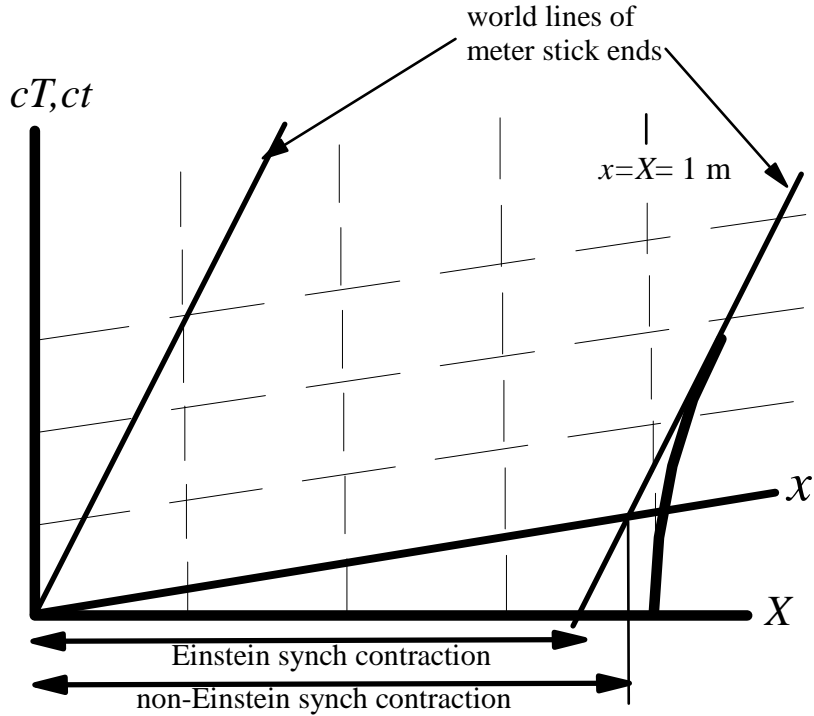}}
\caption{Contraction in Non-Einstein Synched Coordinate System}
\label{fig7}
\end{figure}

Note, from Fig. 2 and Fig. 3, that the eyes of any particular observer will 
see the same thing as an object moves by him, regardless of the 
synchronization convention chosen (i.e., of the slope of the $X$ axis.) The 
only things that matter, as far as visual observation is concerned, are 
where the observer is located and the slope of the world lines of points in 
the object.

\subsection{Conclusion: Measurement of Length}
\label{subsec:mylabel2}
So, consistent with many other analyses of conventionality in 
synchronization, no physical world effect occurs from which one can select 
one synchronization scheme as more fundamental than another. And, the famed 
moving rod ``contraction'' effect varies with synchronization choice.

\section{Revisiting the ``Seeing'' of Fitzgerald-Lorentz Contraction}
\label{sec:revisiting}
Sherwin\cite{Sherwin:1961} describes a means to correct for the speed of 
light using a pulsed radar system, so that the map, plan-position display 
for such a system would show what one would see if one could see the rod 
endpoints instantaneously, rather than time delayed by the light travel 
times. In contemporary times, one could envisage a computer program that 
could also do this, and display an image of the ``true'' Fitzgerald-Lorentz 
contracted meter stick.

However, these approaches determine time delays by assuming a one-way light 
speed of $c$, i.e., Einstein synched observer coordinates. With an alternative, 
equally valid, observer synchronization scheme, and concomitant different 
one-way light speeds, one would see an image with a different length for the 
moving meter stick. So, there really is no ``true'' Fitzgerald-Lorentz 
contracted length factor.

\begin{figure}
\centerline{\includegraphics{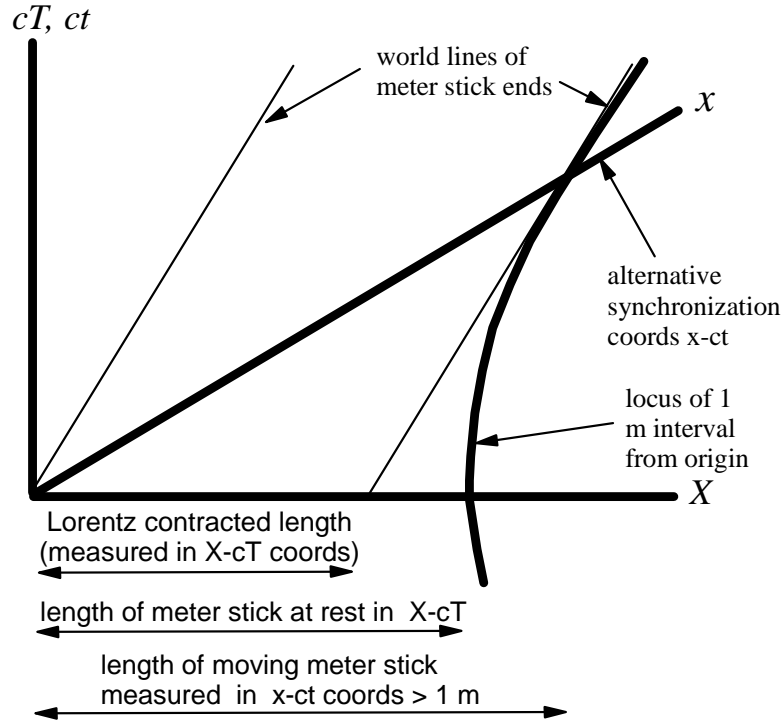}}
\caption{Elongation Measured in Non-Einstein Synched Coordinate System}
\label{fig8}
\end{figure}

\section{The Dewan-Beran-Bell Spaceships}
\label{sec:mylabel1}
Dewan\cite{Dewan:1959}\cite{Dewan:1963}, Beran\cite{Dewan:1959}, and 
Bell\cite{Bell:1987} describe what has become a classic thought experiment 
in favor of the reality of Lorentz contraction. In it, two spaceships, 
S$_{1}$ and S$_{2}$, are connected by a taut thread, and for a period of 
time, carry out identical acceleration histories, as measured by a 
``stationary'' observer. That is, as measured by the stationary observer, 
the two ships increase speed until they reach a constant common velocity, 
all the while keeping the same distance between them. 

From the point of view of Fitzgerald-Lorentz contraction, the thread tries 
to contract, but the motion of the two ships prevents it. Stresses form and 
increase as the ships' speed increases, until the thread breaks. Though many 
of Bell's CERN colleagues originally thought the thread would not break, it 
is now universally agreed that it, indeed, will. One might then presume that 
this generally accepted result ``proves'' the physical reality of 
Fitzgerald-Lorentz contraction.

However, the problem, as stated, presupposes Einstein synchronization. 
Consider the same experiment with different synchronization in the 
stationary observer's frame, such that for the final constant velocity of 
the two ships, there would be no contraction measured on a moving, free 
thread/meter stick. This is a synchronization choice precisely between the 
two regimes wherein a meter stick moving at the final velocity would be 
measured to contract or elongate. While one might naively expect no broken 
thread in this case, one would find the exact same result, a broken thread, 
though the reason is now different. With this synchronization, the lead ship 
S$_{1}$ would be measured to start accelerating before the trailing ship 
S$_{2}$. Thus, the distance between them would not be measured in the 
stationary frame as constant, but increasing. And the increasing distance 
would break the thread.

Any other synchronization scheme would produce the same effect, though the 
degree of length change and difference in starting times would vary. Once 
again, we find the same measurable physical effects, though the conventional 
quantities vary, and there is no way to determine a preferred 
synchronization. The resulting stresses are physically real. But since 
Fitzgerald-Lorentz contraction varies with convention, we would be hard put 
to say that it is physically real.

\section{Time Dilation}
\label{sec:mylabel3}
Note that, like length contraction, we cannot generally ``see'' time dilated 
by the Fitzgerald-Lorentz factor either, as an approaching clock (atomic or 
otherwise) will appear to beat faster than a receding clock, due to the 
Doppler effect. Subtracting this effect out of an observation will, of 
course, give us the appropriate time dilation result. But we could not see 
the effect directly with our eyes.

Nevertheless, if a clock were to be passing by us transversely (direction of 
travel perpendicular to our line of sight), we would see, at least to high 
approximation, a pure time dilation by 1/$\gamma $. Hasselkamp et 
al\cite{Hasselkamp:1979} seems to be the only inertial experiment to date 
that has verified this effect for a detector actually \textit{aimed} at 90 degrees to the 
object's velocity.

With regard to certain means of measurement, however, we run into a similar 
difficulty as with length contraction. If a clock moves from 3D point p to 
3D point q in our frame, we can compare the time on that clock with the 
local clocks at p and q as it passes by them. But the local clock settings 
depend on our choice of synchronization. So, in this regard, a measurement 
of this type for time dilation on the moving clock with respect to time in 
our system must also be ambiguous. 

However, round trip clock time, like round trip light speed, has no 
ambiguity associated with it, as we measure the departure and arrival times 
of the moving clock with the same local clock. And experimentally, this has 
always shown the traveling clock time to be dilated by 1/$\gamma $.

Time dilation and length contraction are further similar in that viewers of 
both located half way between the sides/ends of a moving clock or meter 
stick will find time and length to be altered in the familiar way by a 
factor of $\gamma $.

\section{Summary and Conclusions}
The question posed in the title of this article must be answered with care. 
As seen by the eyes, a moving meter stick does appear to have length unequal 
to one meter. But, due to the finite speed of light, it is not generally 
different by the factor 1/$\gamma = \sqrt {1-v^2/c^2}$, the length looks different to different 
observers, the length can appear greater than 1m, and the length appears to 
change as the object moves by. However, at the specific instant when an 
observer sees himself as midway between the meter stick endpoints, he will 
see a length of 1/$\gamma $ m.

Further, Fitzgerald-Lorentz contraction cannot be measured, in the sense 
that the specific factor by which an object can be considered physically 
contracted depends on the choice of synchronization.  In fact, for some 
synchronization conventions, the object would be measured as \textit{elongated}. The famous 
contraction factor of 1/$\gamma$ is only valid for one 
synchronization convention (Einstein) and thus has no independent physical 
reality.

We must therefore be cautious in 1) promoting time dilation and 
Fitzgerald-Lorentz contraction as similar phenomena that have both been 
verified experimentally, and 2) attempting to extrapolate Fitzgerald-Lorentz 
contraction by assuming it applies directly to certain physical situations. 
The latter include the well known traditional approach to relativistic 
rotation, in which it is often assumed that rods on the circumference of a 
rotating disk contract physically by a factor of 1/$\gamma $. Certain extant 
controversies in relativistic rotation (Rizzi and 
Ruggiero\cite{Rizzi:2004}) may have their roots in this 
assumption\cite{Many:1} (Klauber\cite{Klauber:2007}).

Fitzgerald-Lorentz contraction is unlike round trip time dilation, 
mass-energy dependence, and two-way light speed invariance, which are 
manifestly independent of convention. It is also unlike visual observation 
of moving objects, which is also convention independent. In contrast, it is 
much like, and closely linked to, one-way light speed and synchronization 
gauge.\cite{Readers:2007}

\section{Acknowledgements}
\label{sec:acknowledgements}
I thank Alan Macdonald, John Mallinckrodt, and Robert Krotkov for helpful 
comments, and Alan Macdonald and John Mallinckrodt for bringing to my 
attention certain references with which I was unfamiliar.

-----------------------------------------------


\begin{thebibliography}{30}
\bibitem{Many:1} Advocates of the ``traditional'' approach to relativistic rotation, in which Fitzgerald-Lorentz contraction of the rim of a rotating disk is taken to be a physical reality determining intrinsic surface curvature, sometimes also presume conventionality of simultaneity in rotating frames. It seems one can have one or the other, but not both.
\end{thebibliography}
\end{document}